%% file: routing.tex
\documentclass{llncs}
\usepackage[dvips]{graphicx}
\input{include-no-color.tex}
\newcommand{\lea}{\leftarrow}

\newcommand{\Sig}{\mathsf{S}}
\newcommand{\Lea}{\Leftarrow}

\begin{document}

\title{A Secure Wireless Routing Protocol Using Enhanced Chain Signatures}
\author{Amitabh Saxena}
\institute{International University, Bruchsal 76646, Germany}
\maketitle

\textbf{Abstract:} We propose a routing protocol for wireless networks. Wireless routing protocols allow hosts within a network to have some knowledge of the topology in order to know when to forward a packet (via broadcast) and when to drop it. Since a routing protocol forms the backbone of a network, it is a lucrative target for many attacks, all of which attempt to disrupt network traffic by corrupting routing tables of neighboring routers using false updates. Secure routing protocols designed for wired networks (such as S-BGP) are not scalable in an ad-hoc wireless environment because of two main drawbacks: (1) the need to maintain knowledge about all immediate neighbors (which requires a discovery protocol), and (2) the need to transmit the same update several times, one for each neighbor. Although information about neighbors is readily available in a fairly static and wired network, such information is often not updated or available in an ad-hoc wireless network with mobile devices. Our protocol is a variant of S-BGP called SS-BGP and allows a single broadcast for routing updates without having the need to be aware of every neighboring router. The protocol is based on a novel authentication primitive called Enhanced Chain Signatures (ECS).

\section{Introduction}

The Border Gateway Protocol (BGP)~\cite{RFC4271,rexford02bgp} is a \textbf{Path Vector Routing} protocol, in which routers repeatedly advertise `better' routes (along with the path details) to their immediate neighbors. On receiving an update, a router checks its routing table to decide if this advertised route is better than its existing routes. If so, the router updates its table and advertises the new route to all its other immediate neighbors. The `textbook' variant of BGP (hereafter called BGP) has many security vulnerabilities~\cite{RFC4272,mahajan02understanding}. For instance, a rogue router could claim a shorter route to some destination in order to intercept traffic. Therefore, real implementations use a modified variant of BGP called Secure-BGP (S-BGP). In S-BGP, routers must have knowledge of immediate neighbors and updates are peer-specific. In the context of ad-hoc wireless networks, a node with several receivers in its vicinity must first establish the identity of every such receiver who is also a forwarder, and then broadcast as many updates. Such control plane traffic becomes a bottleneck in scenarios where the wireless devices are densely distributed, power constrained and have low data plane traffic.  

In~\cite{saxena05sig}, a novel signature scheme called Chain Signatures (CS) is proposed. As an application, a secure routing protocol called Stateless Secure BGP (SS-BGP) is also presented. The attack scenario described is of a ``route truncation attack''\footnote{Referred to as a path extraction attack in~\cite{saxena05sig}} in wired networks. SS-BGP is as secure as S-BGP and has some advantages. The main advantages in their protocol over S-BGP are: (1) updates can be broadcast and need not be peer-specific, and (2) routers need not be aware of their immediate neighbors. However, such advantages are not overwhelming in the scenario presented in~\cite{saxena05sig} because true broadcast channels do not exist in wired networks. On the other hand, wireless networks provide true broadcast channels without the ability to control or determine who receives this broadcast. This feature presents a perfect application scenario for SS-BGP. We extend the work of~\cite{saxena05sig} and propose a protocol for wireless routing. The protocol optimizes traffic in the control plane by allowing an ad-hoc network of wireless nodes to establish routing information in presence of several compromised nodes and  without any prior knowledge of topology. The protocol is based on an extension of CS called Enhanced Chain Signatures (ECS). 
\section{Wireless Routing using BGP}

\begin{figure}[htbp]
	\centering
		\includegraphics[width=0.8\textwidth]{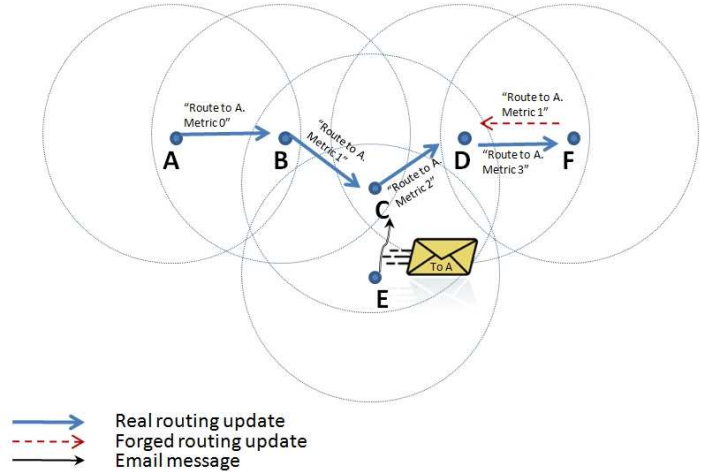}
	\caption{A typical scenario for a route truncation attack in wireless networks.}
	\label{fig:wifi}
\end{figure}

\textbf{Notation:} The following discussion is based on Figure~\ref{fig:wifi}, which shows a wireless network. The circles represent areas of coverage of the transmitter nodes located at their centers, which are represented by small colored discs. The arrows represent various messages broadcast by the nodes at the tail. Note that although the arrows point to particular directions, every node within the corresponding circle is able to receive that message. Each circle has the same radius so that any node $X$ is covered by some node $Y$ iff $Y$ is covered by $X$. Two nodes with non-overlapping coverage can communicate by using intermediate nodes as forwarders. Each node has a permanently active receiver and a passive transmitter that activates when a message is to be sent or a received message is to be forwarded. All messages sent by the transmitter are broadcast to anyone in the covered area. Senders of broadcasts are uniquely determined via this public key. In other words, it is not possible for a broadcasting node to conceal its identity. 

The symbol $X\Lea Y$ denotes the string ``\emph{There is a metric 1 path from $Y$ to $X$}'' and the symbol $X\Lea$ denotes the string ``\emph{There is a metric 0 path from $X$ to $X$}''. The symbol $X\rightarrow: m$ indicates that $m$ is broadcast by $X$. $\Sig_X(m)$ indicates a signature on message $m$ by $X$ using an existentially unforgeable signature scheme (we assume that the signature scheme provides message recovery).

\label{bgp-routing}

\textbf{BGP updates:} (control plane) Refer to Fig.~\ref{fig:wifi}. The following updates are sent for routes to $A$. Each signature (except the first) implies a hop of metric 1.
{\begin{enumerate}
	\item $A \rightarrow:$ $\Sig_A(A\Lea)$
	\item $B \rightarrow:$ $\Sig_A(A\Lea),~\Sig_B(A\Lea B)$
	\item $C \rightarrow:$ $\Sig_A(A\Lea),~\Sig_B(A\Lea B),~ \Sig_C(B\Lea C)$
	\item $D \rightarrow:$ $\Sig_A(A\Lea),~\Sig_B(A\Lea B),~ \Sig_C(B\Lea C),  ~ \Sig_D(  C\Lea D  )$
	\item $E \rightarrow:$ $\Sig_A(A\Lea),~\Sig_B(A\Lea B),~ \Sig_C(B\Lea C),  ~ \Sig_E(  C\Lea E  )$
	\item $F \rightarrow:$ $\Sig_A(A\Lea),~\Sig_B(A\Lea B),~ \Sig_C(B\Lea C),  ~ \Sig_D(  C\Lea D  ),~ \Sig_F(  D\Lea F)$
\end{enumerate}
}

After the above updates have been transmitted and all signatures are verified, each node updates its routing table to contain a tuple (destination, next-hop, metric). For instance $C$'s table will contain the entry $(A, B, 2)$. See~\cite{RFC4271} and Section~\ref{bgp} for details on how the nodes construct the table. The nodes use this table to decide when to broadcast a received data packet and when to keep silent. If a packet arrives from a node that is the next hop for the destination, then the packet is dropped, otherwise it is forwarded.

\textbf{Forwarding:} (data plane) 
 node $E$ broadcasts a data packet destined to $A$. On receiving this packet, node $C$ will activate its transmitter to forward the message (since $C$'s routing table shows that $E$ is not the next-hop on route to $A$). Both $B$ and $D$ will receive this broadcast from $C$ but only $B$ will activate to forward it further (since $D$'s routing table shows that $C$ is the next-hop for destination $A$). Finally, the broadcast of $B$ is received by $A$ and there is no further forwarding.

\textbf{Route truncation attack:} 
\label{route-atk}
Although the signatures ensure that fake routes cannot be created, they do not ensure that intermediate routes are not truncated. As an example, $F$ is an attacker who needs to intercept the above data packet. First note that using the above updates, $D$'s routing table is set to discard data packets received from $C$ and addressed to $A$. Thus, such packets would never be received by $F$. To launch its attack, $F$ pretends to have a shorter route to $A$ than $C$ does. To do this, $F$ replaces its routing update broadcast in Step 6 with the following:

$F \rightarrow: \Sig_A(A\Lea),~\Sig_F(A\Lea F)$

On receiving this update, node $D$ will believe that the route to $A$ via $F$ is shorter. Consequently, $D$ will update its routing table to forward a data packet received from $C$ and addressed to $A$. This general attack is called \emph{route truncation}. In this attack, every data packet sent by $E$ and addressed to $A$ will be received by $F$. Although we only considered an eavesdropping attack,  it is trivial for $F$ to launch a DoS attack. For instance, if $F$ drops all data plane traffic, it can ensure that data packets originating from $D$ and addressed to $A$ never reach their destination.

\textbf{Secure-BGP Routing:} A possible way to disallow this attack is to use Secure-BGP (S-BGP), which is an augmented version of BGP. S-BGP requires that updates be recipient specific. Although S-BGP is designed for wired networks, the same concept can be adapted to wireless. S-BGP requires that each host be aware of their immediate neighbors (in this context, receivers within its coverage). S-BGP assumes that this information has somehow been established. Let $X$ and $Y$ denote nodes within $E$'s and $F$'s coverages respectively (but not covered by any other node). The S-BGP updates are as follows.

\begin{enumerate}
	\item $A \rightarrow:$ $\Sig_A(A\Lea B)$
	\item $B \rightarrow:$ $\Sig_A(A\Lea B),~\Sig_B(B\Lea C)$
	\item $C \rightarrow:$ $\Sig_A(A\Lea B),~\Sig_B(B\Lea C),~ \Sig_C(C\Lea D)$
	\item[]$C\rightarrow:$ $\Sig_A(A\Lea B),~\Sig_B(B\Lea C),~ \Sig_C(C\Lea E)$
	\item $D \rightarrow:$ $\Sig_A(A\Lea B),~\Sig_B(B\Lea C),~ \Sig_C(C\Lea D),  ~ \Sig_D(  D\Lea F  )$
	\item $E \rightarrow:$ $\Sig_A(A\Lea B),~\Sig_B(B\Lea C),~ \Sig_C(C\Lea E),  ~ \Sig_E(  E\Lea X  )$
	\item $F \rightarrow:$ $\Sig_A(A\Lea B),~\Sig_B(B\Lea C),~ \Sig_C(C\Lea D),  ~ \Sig_D(  D\Lea F  ),~ \Sig_F(  F\Lea Y  )$
\end{enumerate}

The above protocol is secure from route truncation attack. From an application perspective, the only difference between (ordinary) BGP and S-BGP is that while BGP is resistant to every attack except route truncation attacks, S-BGP is also resistant to such attacks.

\textbf{Stateless Routing:} Observe that the S-BGP protocol of Example 2 has two major drawbacks: (1) Each router must be ``aware'' of its neighbors, and (2) In the example, router $C$ can no longer broadcast the same message for every neighbor. This has scalability problems as follows. Firstly, every transmitter must have prior knowledge of all receivers within its coverage, which is clearly problematic. Secondly, since each update is peer-specific, even a single route change could result in a large number of broadcasts by a node with many receivers in its coverage. 
It would be much simpler if the underlying routing protocol resisted route truncation attacks and required each router to broadcast only one short message on each update without being aware of its neighbors/receivers. We call such a protocol a \textbf{Stateless Routing Protocol}. To avoid the route truncation attack in a stateless protocol, given the message in Step 4 of Example 1, attacker $F$ should not be able to extract $\Sig_A(A\Lea)$. 
\textbf{Our Contribution:} We present a stateless routing protocol that resists route truncation attacks. Our proposed protocol, called Stateless Secure-BGP (SS-BGP) is a variation of S-BGP and provides the following benefits:
\begin{enumerate}
	\item It is fully stateless - routers need not be aware of their neighboring receivers.
	\item It is communication efficient - one constant size broadcast per update irrespective of the number of peers. 
\end{enumerate}

\subsection{Related Work} 
\label{related}
Current research assumes the stateful scenario of Example 2 (S-BGP), and is focused on reducing the number of signatures transmitted and/or processing time~\cite{boneh03aggregate,park02efficient}. For instance, aggregate signatures have been proposed to keep the signature payload to a constant size~\cite{boneh03aggregate}. The authors of~\cite{zhao05aggregated} propose the use of \emph{Signature Amortization}~\cite{park02efficient} coupled with aggregate or sequential aggregate signatures~\cite{anna03sequential} to reduce the size of update messages and the signing time. The authors of~\cite{OMSgentry} propose the use of identity-based sequential aggregate signatures (IBSAS) to authenticate routing updates. However, the above works assume a stateful environment, where information about immediate peers is known. and do not consider the route truncation attack described above using Example 1 (where information about the next-hop is not available to the current signer). Specifically, the above works do not consider the attack where given the message in Step 6, router $F$ is able to compute the message sent in Step 1 without extracting the private keys of all of $\{A, B, C, D, E\}$.

\section{The Building Blocks}

\textbf{Notation:} We first develop some notation to deal with ordered elements, which we call sequences. 

\begin{enumerate}
	\item A \emph{sequence} is similar to a set except that the order of its elements is important. Elements of a sequence are written in order, and enclosed within the symbols $\la, \ra$. For instance, $\la y_1, y_2, y_3\ra$ is a sequence. The symbol $\theta$ denotes the empty sequence with zero elements. 	\item Let $\ell_a=\la y_1, y_2,\ldots, y_k\ra$ be some non-empty sequence. For any other sequence $\ell_b$, we say that $\ell_b\prec \ell_a$ if and only if $\ell_b=\la y_1, y_2,\ldots, y_i\ra$ and $0\leq i \leq k$. We say that two sequences $\{\ell_a, \ell_b\}$ \textbf{overlap} if there exists a non-empty sequence $\ell$ such that $\ell\prec \ell_a$ and $\ell\prec \ell_b$. For instance, $\{\la y_1, y_2\ra, \la y_1\ra\}$ overlap, while $\{\la y_1, y_2\ra,\la y_2\ra\}$ do not.
	\item For any two sequences $\ell_a, \ell_b$, the symbol $\ell_a\cup \ell_b$ denotes the \textbf{set} of elements that belong to at least one of $\{\ell_a, \ell_b\}$. Similarly $\ell_a\cap \ell_b$ denotes the \textbf{set} of elements that belong to both $\ell_a$ and $\ell_b$. We denote by $\ell_a\odot \ell_b$ to be the \textbf{set} of elements from the largest sequence $\ell$ such that $\ell\prec \ell_a$ and $\ell\prec \ell_b$. Clearly, for two overlapping sequences $\{\ell_a, \ell_b\}$, we have that $\ell_a\odot \ell_b\neq \emptyset$.
	\item Collapsing Rule: Any sequence $\la \la y_1, y_2, \ldots y_i\ra, y_{i+1}\ra$ is equivalent to the sequence $\la y_1, y_2,\ldots  y_{i+1}\ra$. 
\end{enumerate}

\subsection{Enhanced Chain Signatures}

In~\cite{saxena05sig}, a novel signature scheme called Chain Signatures (CS) is presented, which is essentially a combination of Boneh~\emph{et al.}'s aggregate signatures and Verifiably Encrypted Signatures (VES)~\cite{boneh03aggregate}. The nice property about CS is that in addition to ordinary properties, they also provide truncation resilience. In the model of~\cite{saxena05sig}, every signer signs the same message. We consider an extension of CS, called Enhanced Chain Signatures (ECS) in which each signer may sign different messages. Note that CS can be extended to allow signers to sign different messages by using another existentially unforgeable signature scheme in conjunction as discussed in the original paper. However, ECS offers better performance and a cleaner security definition. ECS is described below.

\paragraph{Algorithms:} ECS are defined by 3 algorithms: \textbf{ECS-}(\textbf{KeyGen}, \textbf{Sign}, \textbf{Verify}).
	\begin{description}
	\item [ECS-KeyGen$(\tau)$] takes a parameter $\tau$. It outputs a private-public key pair $(x, y)$. 	\item [ECS-Verify$(\ell, \sigma)$] takes as input $\ell=\la (m_1,y_1), (m_2,y_2), \ldots\ra$, a finite sequence of (message, public key) pairs, and a string $\sigma$. 		\begin{enumerate}
			\item If $\ell=\theta$ and $\sigma=1$, the algorithm outputs \textsf{VALID}.
			\item If $\ell=\theta$ and $\sigma\neq 1$, the algorithm outputs \textsf{INVALID}.
			\item If any public key $y_i$ repeats in $\ell$, the algorithm outputs \textsf{INVALID}.
			\item If this step is executed, the algorithm invokes a deterministic poly-time procedure after which it outputs either \textsf{VALID} or \textsf{INVALID}. 
		\end{enumerate}
	\item [ECS-Sign$(x_i, y_i, m_i, \ell_j, \sigma_j)$] takes five inputs, which can be grouped into three parts: (1) a (private key, public key, message) tuple $(x_i, y_i, m_i)$, (2) a sequence $\ell_j=\la (m_1,y_1), (m_2,y_2), \ldots, (m_j,y_j)\ra$ of $j$ (message, public key) pairs for $j\geq 0$, and (3) a string $\sigma_j$ (a purported chain signature on $\ell_j$). 	\begin{enumerate}
		\item If $y_i\in\{y_1,y_2,\ldots, y_j\}$, the algorithm outputs \textsf{ERROR}. 
		\item If \textbf{ECS-Verify}$(\ell_j,\sigma_j)=\mathsf{INVALID}$, the algorithm outputs \textsf{ERROR}. 
		\item If this step is executed, the algorithm computes a chain signature $\sigma_i$ on $\ell_i$, where $\ell_i=\la \ell_j, (m_i,y_i)\ra$. It outputs $\sigma_i$.
	 \end{enumerate}
\end{description}

A string $\sigma$ is an ECS signature on a sequence $\ell$ if $\textbf{ECS-Verify}(\ell,\sigma)=\textsf{VALID}$. 

\paragraph{Difference with a CS scheme:} If for every sequence $\la (m_1,y_1), (m_2,y_2), \ldots\ra$ to be signed, holds $\forall i:m_i=m_1$, then the ECS scheme is also a CS scheme. 
\paragraph{Security Model:} We define the security of ECS in the adaptive known key (AKK) model.\footnote{We note that our actual construction is actually secure in the stronger model of \emph{Adaptive Chosen Key (ACK)} attacks, where the adversary inserts/replaces chosen public keys at chosen locations (and hands over the corresponding private keys to the challenger). In the extended model, such public keys are considered equivalent to those that have been extracted. } This notion is called security under \emph{adaptive known key and chosen message attack}. We define two variants using a parameter $\omega$ such that setting $\omega=1$ results in the weaker variant. \begin{description}
	\item []\begin{center} \textbf{Game ECS-UNF$_\omega(\tau)$} \end{center} 
	\item [Setup.] The challenger gives the security parameter $\tau$ to the adversary $\mathcal{A}$, who then selects a game parameter $n$ (the number of public keys) and an $n$ bit string $extr$ (the 1s of $extr$ denote the indexes of the public keys that the adversary wants to extract). Let $extr[i]$ denote the $i^{th}$ bit of $extr$.
	On receiving $(n, extr)$, the challenger generates $n$ key-pairs $(x_i, y_i)\rand$ \textbf{ECS-KeyGen}$(\tau)~(1\leq i \leq n)$ and gives the set $Y=\{y_i\}_{1\leq i\leq n}$ of $n$ public keys along with set $X=\{x_i|extr[i]=1\}_{1\leq i\leq n}$ of extracted private keys to $\mathcal{A}$. 
	
	In the following, we denote by $L$ the set of all non-empty sequences of pairs $(m, y)\in\{0,1\}^*\times Y$ such that no $y$ is repeated. 		\item [Queries.] Working adaptively, $\mathcal{A}$ makes queries as follows:
	\begin{enumerate}
	\item \emph{Extract}: This query consists of a public key $y\in Y$. If $\omega=1$, the challenger responds with $\bot$. Otherwise, it responds with the private key corresponding to $y$. 
	\item \emph{ECS-Sign}: This query consists of a sequence $\ell\in L$. The challenger responds with an ECS signature on $\ell$. 
	\end{enumerate}
	\item [Output.] $\mathcal{A}$ outputs a pair $(\ell_A,\sigma_A)\in L\times \{0,1\}^*$.
	\item [Result.] $\mathcal{A}$ wins if \textbf{ECS-Verify}$(\ell_A,\sigma_A)=\mathsf{VALID}$ and $\ell_A$ is \emph{non-signable} (Def.~\ref{nonsignable}). 
\end{description}

\begin{defn}\label{nonsignable}
(\textbf{Non-signable Sequence}) In the game, let $Y_X\subset Y$ and $L_S\subset L$ be the set of inputs to the extract and ECS-sign queries respectively (Note: $\{y_i|extr[i]=1 \}_{1\leq i \leq n}\subseteq Y_X$). For any $\ell_A \in L$, define the set $L_A=\{\ell_i|\ell_i\in L_S\wedge \{\ell_A,\ell_i\}\mbox{ overlap}\}$. $\ell_A$ is \emph{non-signable} if: 
				\begin{enumerate}
					\item $\ell_A\notin L_S$ 
					\item The set $\{(m_i,y_i)|(m_i,y_i)\in \ell_A\wedge y_i\notin Y_X\}$ is non-empty.
					\item For every $\ell_i\in L_A$, the set $\{(m_j,y_j)|(m_j,y_j)\in (\ell_i\cup \ell_A)\backslash (\ell_i\odot \ell_A)\wedge y_i\notin Y_X\}$ is non-empty.
				\end{enumerate}
\end{defn}

In the following, we also consider the use of hash functions in the construction. 

\begin{defn} \label {def-chain} The ECS scheme $\Sigma$ is $(n, \tau, t, q_s, q_e, q_h, \epsilon)$-$\mathbf{UNF}$-$\omega$-secure for $\omega\in\{1,2\}$ if, for  game parameters $\tau$ and $n$, there is no adversary $\mathcal{A}$ that runs for time at most $t$; makes at most $(q_s, q_e, q_h)$ chain-sign, extract and hash queries respectively; and wins Game ECS-UNF$_\omega(\tau)$ with probability at least $\epsilon$. \end{defn}

\paragraph{Discussion:} Roughly speaking, the above security notion implies security under two types of forgeries~\cite{saxena05sig}. We illustrate this with an example. The first forgery (called \emph{ordinary forgery}) occurs when the adversary manages to output a valid ECS signature on a sequence $\ell=\la(m_1,y_1), (m_2, y_2)\ra$ after making an ECS-sign query on $\ell_1=\la(m_1, y_1)\ra$ but without making an extract query on $y_2$. This is the type of forgery that all multisignatures schemes (including the ones discussed in Section~\ref{related}) resist. The second type of forgery in ECS (called \emph{extraction forgery}) occurs when the adversary manages to output a valid ECS signature on $\ell=\la(m_1,y_1), (m_2, y_2)\ra$ after making an ECS-sign query on $\ell_3=\la(m_1, y_1),(m_2, y_2),(m_3, y_3)\ra$ but without making an extract query on $y_3$ and one of $\{y_1,y_2\}$. A scheme secure against an extraction forgery is said to be \emph{truncation resilient}. Note that none of the schemes mentioned in Section~\ref{related} consider extraction forgery in their security definitions. 

\paragraph{Construction:} A construction of ECS is given in Appendix~\ref{app:cons}.

\subsection{The BGP Routing Protocol} 
\label{bgp}
It is helpful to refer to Figure~\ref{fig:wifi} in the following discussion. As mentioned earlier, this protocol is not resistant to route truncation attacks.

Consider a routing update for a particular destination. This routing update propagates in a structure represented by a tree rooted at the destination node with routers at level $i$ representing nodes of the tree at level $i$ of the update (the root node is considered to be at level 1). For any positive integer $i$, consider an arbitrary node at level $i+1$ of the tree. Label as $R_1, R_2, \ldots, R_{i+1}$, the sequence of nodes in the path from the root node to this node (both inclusive). Denote by \textsf{Sign}$_i$, \textsf{Verify}$_i$ the sign and verify functions of node $R_i$ under an existentially unforgeable signature scheme. $R_0$ is a constant string used for notational convenience. BGP has two phases: Initialize and Update.
\begin{description}  
\item [Initialize] Let $t_1$ be a timestamp when this update is initiated. The initiator ($R_1$) sets $m_1\lea(R_0, R_1, t_1)$; $Sig_1\lea\textsf{Sign}_1(m_1)$; and $U_1\lea\la (m_1, Sig_1)\ra$. It broadcasts $U_1$. 

\item [Update] On receiving update $U_i$, node $R_{i+1}$ sets $m_{i+1}\lea(R_{i}, R_{i+1}, t_{i+1})$, where $t_{i+1}$ is a timestamp when this update was received. The update phase consists of two stages: 
	\begin{enumerate}
		\item \emph{Validation:} Parse $U_i$ as $\la(m_1,sig_1), (m_2,sig_2), \ldots,(m_i, sig_i)\ra$. Then ensure the following: 
			\begin{enumerate}
				\item For each $j\in[1..i]: m_j$ is of the form $(R_{j-1}, R_j, t_j)$. 
				\item For each $j\in[1..i]:R_{j+1}\notin\{R_1,R_2,\ldots, R_j\}$.
				\item The route to $R_1$ given by the sequence $\la R_1, R_2, \ldots, R_i\ra$ is either new or better than an existing route.
				\item For each $j\in[1..i]:$ The difference in timestamps, $t_{j+1}-t_j$ is within a pre-defined threshold $t$.
				\item For each $j\in[1..i]:\textsf{Verify}_j(m_j, sig_j)=\textsf{VALID}$.
\end{enumerate}
			Abort if any of the above checks fail, otherwise, update routing table and proceed to the next step. 
		\item\emph{Propagation:}   Set $sig_{i+1}\lea \textsf{Sign}_{i+1}(m_{i+1})$; $U_{i+1}\lea\la U_i, (m_{i+1},sig_{i+1}))\ra$; and broadcast $U_{i+1}$.
	\end{enumerate}
\end{description}
If the update has been successfully been validated and propagated, then we say that the update has been \emph{accepted}. 
\paragraph{Correctness:} Step 1. ensures that the update is of the correct format. If all nodes are honest then the only steps  where validation can fail are
\begin{enumerate}
	\item Step ii., when $R_{i+1}\in\{R_1,R_2,\ldots, R_i\}$. For instance, referring to Figure~\ref{fig:wifi}, when $C$'s routing update reaches $B$. 
	 
	\item Step iii., when the existing routing table has a better route.
\end{enumerate}
In either case, the protocol behaves correctly and implements BGP~\cite{RFC4271}.

\paragraph{Security:} The security is captured in the Validation stage as follows:	
					\begin{enumerate}
						\item Step iv. prevents replay attacks.
						\item	Step v. ensures each node $j\in[1..i]$ accepted this update. 
					\end{enumerate} 

\paragraph{Weakness:} The weakness in this protocol is that although Step v. ensures that each node $j\in[1..i]$ accepted this update, it does not ensure that these are the only nodes that accepted this update. Specifically, it does not prevent the route truncation attack discussed in Section~\ref{route-atk}. For instance, when $R_{i+1}$ receives this update, it cannot ensure that $R_i$ received this update from $R_{i-1}$, since $R_i$ could have truncated several intermediate entries. Due to this weakness, the above protocol is not suitable for use in an real-world environment.

\paragraph{Advantages:} The primary advantage is that of statelessness - any node $R_i$ never needs knowledge of $R_{i+1}$. Therefore, there is no need to establish knowledge of immediate neighbors in order to use the protocol. Secondly, since update $R_i$ is independent of $R_{i+1}$, the same update can be used by several nodes at level $i+1$.

\section{Stateless Secure BGP using ECS}

\label{ssbgp}

As in Section~\ref{bgp}, let $\la R_1, R_2, \ldots\ra$ be any arbitrary sequence of nodes that would be affected by a given update\footnote{Note that there will be many such distinct sequences for the same update and $R_1$ would be the first node in all these sequences.}, and $t_i$ be the timestamp at which the update originated/arrived at node $i$. Let $(x_i,y_i)$ be the (private, public) key-pair of node $R_i$ under an ECS scheme. Assume that every node has a distinct public key. The SS-BGP protocol is as follows.
\begin{description}
\item [Initialize] The initiator ($R_1$) sets: $m_1\lea t_1$; $L_1\lea \la(m_1, R_1)\ra$; $\ell_1=(m_1,y_1)$; and $\sigma_1\lea \textbf{ECS-Sign}(x_1,y_1, \ell_1, \theta, 1)$. Finally it sets $U_1\lea (L_1, \sigma_1)$ and broadcasts the update $U_1$.\footnote{\label{gps}Our basic BGP variant only needs to validate the timestamp. However, if any additional information must be inserted at node $i$, this can be included as part of $m_i$. A possible example of such additional information is the GPS coordinates of $i$. }	 

\item [Update] On receiving update $U_i=(L_i,\sigma_i)$, node $R_{i+1}$ sets $m_{i+1}\lea t_{i+1}$ and does the following:
	\begin{enumerate}
		\item  \emph{Validation:} Parse $L_i$ as $\la(m_1,R_1), (m_2,R_2), \ldots,(m_i, R_i)\ra$. Then ensure the following: 
			\begin{enumerate}
				\item For each $j\in[1..i]:m_j$ is of the form $t_{j}$. 
				\item For each $j\in[1..i]:R_{j+1}\notin\{R_1,R_2,\ldots, R_j\}$.
				\item The route to $R_1$ given by the sequence $\la R_1, R_2, \ldots, R_i\ra$ is either new or better than an existing route.
				\item For each $j\in[1..i]:$ The difference in timestamps, $t_{j+1}-t_j$ is within the pre-defined threshold $t$.
 				\item Construct the sequence $\ell_i=\la(m_1, y_1), (m_2, y_2), \ldots, (m_i,y_i)\ra$ and test if $\textbf{ECS-Verify}(\ell_i, \sigma_i)=\textsf{VALID}$. 
			\end{enumerate}
			Abort if any of the above checks fail, otherwise, update routing table and proceed to the next step. 
		\item \emph{Propagation:} Set $L_{i+1}\lea \la L_i,(m_{i+1},R_{i+1})\ra$; $\ell_{i+1}\lea \la \ell_i,(m_{i+1},y_{i+1})\ra$; $\sigma_{i+1}\lea\textbf{ECS-Sign}(x_{i+1},y_{i+1}, m_{i+1}, \ell_i,\sigma_i)$. Finally, set $U_{i+1}\lea(L_{i+1}, \sigma_{i+1})$ and broadcast $U_{i+1}$.
	\end{enumerate}
\end{description}

\paragraph{Correctness:} Referring to the basic BGP protocol of Section~\ref{bgp}, the only difference is in Step v. Assuming that the validation in this step always passes, the above protocol then implements BGP in presence of honest users. 

\paragraph{Security}: The only difference from the protocol of Section~\ref{bgp} regarding security is in Step v., where \textbf{ECS-Verify} is used. First note that if the ECS scheme is \textbf{UNF-1} secure then the protocol indeed ensures that each node $j\in[1..i]$ accepted this update. Now consider Figure~\ref{fig:wifi}. Then the nodes in the path of the update received by $F$ for destination $A$ are $(A, B, C, D)$. The update is of the form $U_D=(L_D,\sigma_D)$, where $L_D=\la(t_A, A)(t_B, B)(t_C, C)(t_D, D)\ra$. Consider the only two possible attacks by $F$. 

\begin{enumerate}
	\item \textbf{Route Truncation Attack:} As an example, consider the attack of Section~\ref{route-atk}, where given the update \[\left[\Sig_A(A\Lea),~\Sig_B(A\Lea B),~ \Sig_C(B\Lea C),  ~ \Sig_D(  C\Lea D  )\right],\] attacker $F$ extracts the update $\left[\Sig_A(A\Lea)\right]$. In SS-BGP, this translates to $F$ extracting the ECS signature $\sigma_A$ on $\ell_A$ given the ECS signature $\sigma_D$ on $\ell_D$. Under the \textbf{UNF-1} security notion of ECS, this is not possible unless $F$ has extracted all the private keys of $\{B, C, D\}$.
	\item \textbf{Repeating Attack:} $F$ simply repeats (forwards) the message received from $D$ and masquerades as $D$ in the hope of making the route at least one hop shorter. This attack is unavoidable using all existing methods for wired networks, including those based on S-BGP and the ones discussed in~\ref{related}. An attacking node can always forward messages without modification so that its presence is never detected. However, in the case of wireless networks, there are possible ways to detect this attack. 
	
	Under our original assumption that senders of broadcasts can be uniquely identified, the repeating attack is not possible. Note that this assumption is quite strong. A possible way to avoid this assumption would be to encode the GPS coordinates of each node in the update along with the timestamp (see Footnote~\ref{gps}). Using this technique, it is possible to determine if a message was sent via a repeater or not. \end{enumerate}

\paragraph{Communication Overhead}: Assuming that public keys can be uniquely identified by IP addresses, the sequences $\ell_i$ can be constructed at the receiver's end by knowing the timestamps $(t_1, t_2, \ldots, t_i)$ and IP addresses $(R_1, R_2,\ldots, R_i)$. Consequently, the only overhead in this protocol is that of one single chain signature $\sigma_i$, which is an element of $G_1$ and is less than 30 bytes using the parameters of~\cite{saxena05sig}. Contrast this with basic BGP or S-BGP, both of which incur an overhead of $i$ signatures. 
\paragraph{Storage and Performance:} The keys are elements of $G_1$ and can be stored in $\leq 30$ bytes~\cite{saxena05sig}. The benchmarks of~\cite{ateniese05improved} indicate the following performance estimates of the above protocol (assuming $n$ nodes in the path):
  	\begin{enumerate}
  		\item \textbf{Update Propagation}: one exponentiation and addition in $G_1$, and one computation of $\mathcal{H}$ (total $< 2$ms).
  		\item \textbf{Update Verification}: $n$ pairing computations and multiplications in $G_2$, and $n$ computations of $\mathcal{H}$ (giving $< 1$ second for $n=100$).
  	\end{enumerate}

 To conclude, SS-BGP based on ECS is at least as secure and as computationally efficient as S-BGP based on the signature schemes of~\cite{boneh03aggregate,OMSgentry} without the extra overhead of neighbor discovery, multiple signature computation, multiple broadcasts and multiple signatures in each broadcast.

\paragraph{Multiple Updates Aggregation}: In the above description, we assumed that each advertisement $U_i$ contains only one route and is transmitted instantaneously. In the real world, each advertisement contains multiple routes and is sent periodically. Fortunately, the ECS scheme allows for \emph{signature aggregation} and \emph{aggregate verification} where a large number of ECS signatures are verified at once~\cite{boneh03aggregate}.

\section{Conclusion}

In this paper we presented an efficient stateless routing protocol for wireless networks called Stateless Secure BGP (SS-BGP). Our protocol is based on ordinary BGP, which allows forwarding nodes to build a routing table without prior knowledge of neighboring forwarders, and allowing a single broadcast per node irrespective of the number of neighbors. We call such a protocol \emph{stateless}. However, BGP used in this manner does not resist route truncation attacks, and is therefore useless for practical purposes. In order to prevent such attacks, modern implementations use a stateful variant of BGP called Secure-BGP (S-BGP). S-BGP requires forwarding nodes to have prior knowledge of their immediate neighbors and requires as many broadcasts per node as there are neighbors. Although sufficient for wired networks, S-BGP has scalability issues in mobile ad-hoc wireless networks with low data plane traffic because of the need to constantly keep updated information about neighbors in order to keep updated routing tables. 

SS-BGP is based on stateless BGP and resists route truncation attacks. The main ingredient of our protocol is an authentication primitive called Enhanced Chain Signatures (ECS), which is an extension of Chain Signatures (CS) of~\cite{saxena05sig}. The main property of CS and ECS that differentiates them from other multisignature schemes is that in addition to standard authentication, both primitives also provide \emph{truncation resilience} (see~\cite{saxena05sig} for a formal discussion of this concept). In our context, this property translates to route truncation resilience in a stateless implementation of BGP. 

In summary, S-BGP incurs the following overhead: (1) neighbor discovery, (2) multiple signature computation, (3) multiple broadcasts, and (4) multiple signatures in each broadcast. Although the signature schemes of~\cite{boneh03aggregate,OMSgentry} are able to address issue (4), they 
do not address the remaining three. SS-BGP based on ECS is as secure and computationally efficient as S-BGP based on the signature schemes of~\cite{boneh03aggregate,OMSgentry} without any of the above overheads. This feature makes SS-BGP particularly suited for use in a wireless environment. \appendix

\bibliographystyle{unsrt}
\bibliography{amitbib}

\section{Construction of ECS}
\label{app:cons}
The following construction of ECS is an extension of the CS scheme of~\cite{saxena05sig}.
\begin{enumerate}
	\item \textbf{Bilinear Maps:} Let $G_1$ and $G_2$ be two cyclic multiplicative groups both of prime order $q$ such that computing discrete logarithms in $G_1$ and $G_2$ is intractable. A bilinear pairing is a map $\hat{e} : G_1 \times G_1 \mapsto G_2$ that satisfies the following properties~\cite{boneh04short,boneh01identitybased,boneh03aggregate}.\begin{enumerate}
\item	\emph{Bilinearity}: $\hat{e}(a^x, b^y) = \hat{e}(a, b)^{xy}~\forall a, b \in G_1$ and $x, y \in \mathbb{Z}_q$.
\item \emph{Non-degeneracy}: If $g$ is a generator of $G_1$ then $\hat{e}(g, g)$ is a generator of $G_2$.
\item \emph{Computability}: The map $\hat{e}$ is efficiently computable.
\end{enumerate}
In a practical implementation, $G_1$ is a subgroup of the (additive) group of points on the elliptic curve and $G_2$ is the multiplicative subgroup of a finite field. The map $\hat{e}$ is derived either from the modified Weil pairing~\cite{boneh04short,boneh01identitybased} or the Tate pairing~\cite{paulo02efficient}. The security of our protocol depends on the hardness of the Computational Diffie-Hellman (CDH) problem in $G_1$, defined as follows: given $g^x, g^y\in G_1$ for some generator $g$ and unknowns $x,y$, compute $g^{xy}\in G_1$~\cite{boneh03aggregate}.	\item \textbf{Common Parameters:} Let $\he: G_1 \times G_1 \mapsto G_2$ be a bilinear map over cyclic multiplicative groups $(G_1, G_2)$ of prime order $q$ and $g\in G_1$ be a generator of $G_1$. The prime $q$ is chosen so that the CDH problem in $G_1$ is requires $\approx 2^\tau$ operations for some security parameter $\tau$. See~\cite{boneh03aggregate} for details. Let  $\mathcal{H}:\{0,1\}^*\mapsto G_1$ be a hash function. In the rest of this section, these parameters will be considered common and public.
\item \textbf{The Algorithms:} The following are the three algorithms in the scheme. 
\begin{description}
	\item [ECS-KeyGen.] Each user $j$ generates $x_j\rand\mathbb{Z}_q$. The private key of $j$ is $x_j$. The corresponding public key is $Y_j=g^{x_j}\in G_1$.  
	\item [ECS-Sign.] Let $\ell_i=\la (m_1,Y_1), (m_2,Y_2), \ldots, (m_i,Y_i)\ra$ be some sequence of $i$ distinct (message, public key) pairs. An ECS signature on sequence $\ell_i$ is an element $\sigma_i\in G_1$, where:
\[\sigma_i = \prod_{j=1}^{i}\mathcal{H}(\la (m_1,Y_1), (m_2,Y_2),\ldots, (m_j,Y_j)\ra)^{x_j},\]
 and $\sigma_0=1_{G_1}$. The ECS signature $\sigma_i=\textbf{ECS-Sign}(x_i,y_i,m_i,\ell_{i-1},\sigma_{i-1})$ is computed by user $i\in\{1,2,\ldots\}$ as \[\sigma_i\leftarrow\sigma_{i-1}\cdot\mathcal{H}(\la (m_1,Y_1), (m_2,Y_2),\ldots, (m_i,Y_i)\ra)^{x_i}.\]
	\item [ECS-Verify($\ell_i,\sigma_i$).] To verify the signature $\sigma_i$ on $\ell_i=\la (m_1,Y_1), (m_2,Y_2),\ldots, (m_j,Y_j)\ra$, check that all $Y_i$'s are distinct and the following holds:
\[\he (\sigma_i, g) \stackrel{?}{=} \prod_{j=1}^{i}\he(\mathcal{H}(\la (m_1,Y_1), (m_2,Y_2),\ldots, (m_j,Y_j)\ra), Y_j).\]
\end{description}

\item \emph{Security:} The CS scheme of~\cite{saxena05sig} is shown to be $\mathbf{UNF}$-$1$-secure. We observe that the same proof of~\cite{saxena05sig} works for the above ECS construction without any modification in the simulator. Hence, the above ECS construction is also $\mathbf{UNF}$-1-secure in the random oracle model under the computational Diffie-Hellman (CDH) assumption in bilinear maps. We refer the reader to~\cite{saxena05sig} for the original proof. A sketch will be given in the full version of this paper on a preprint archive.
\end{enumerate}
\end{document}

%% file: include-no-color.tex
\usepackage[all]{xy}
\usepackage{amsfonts}
\usepackage{amsmath}
\usepackage{amssymb}
\newtheorem{thm}{Theorem}[section]
\newtheorem{defn}[thm]{Definition}

\newcommand{\rand}{\stackrel{R}{\leftarrow}}	
\newcommand{\la}{\left\langle}
\newcommand{\ra}{\right\rangle}

\newcommand{\he}{\hat{e}}